\documentclass{aa501}
\usepackage[]{natbib}
\usepackage[]{graphics}
\bibpunct{(}{)}{;}{a}{}{,}
\newcommand{\etal}{{\it et al.}\ }
\newcommand{\Msun}{\hbox{M$_\odot$}\ }
\newcommand{\sbu}{\mbox{mag arcsec$^{-2}$}\ }
\newcommand{\kms}{\mbox{km~s$^{-1}$}\ }
\begin{document}

\title{The visible environment of galaxies with counterrotation }

\author{D. Bettoni\inst{1} 
\and
G. Galletta\inst{2}
\and
F. Prada\inst{3}}

\institute{Osservatorio Astronomico di Padova, Vicolo dell'Osservatorio 5, 35122 Padova, Italy
\and 
Dipartimento di Astronomia, Universit\`a di Padova, Vicolo dell'Osservatorio 2, 35122 Padova, Italy
\and
Centro Astron\'omico Hispano-Alem\'an, Apartado 511, E-04080 Almer\'\i a, Spain
}

\offprints{G. Galletta}

\authorrunning{Bettoni \etal}
\titlerunning{Environment of counterrotation}

\date{Received 6 February 2001; accepted 23 April 2001}

\abstract{    
In this paper we present a statistical study of the environments of 49
galaxies in which there is gas- or stellar- counterrotation.  The
number of possible companions in the field (to apparent magnitude 22),
their size and concentration were considered. All the statistical
parameters were analysed by means of Kolgomorov-Smirnov tests, 
using a control sample of 43 galaxies without
counterrotation. From our data, no significant differences between the
counter-rotating and control samples appear.  This is different to
Seyfert or radio-loud galaxies which lie in environments with a higher
density of companions. On the contrary, if a weak tendency exists,
for galaxies with gas counterrotation only, it is discovered in
regions of space where the large scale density of galaxies is
smaller. \\ Our results tend to disprove the hypothesis that
counterrotation and polar rings derive from a recent interaction with
a small satellite or a galaxy of similar size. To a first
approximation, they seem to follow the idea that all galaxies are
born through a merger process of smaller objects occurring very
early in their life, or that they derive from a continuous, 
non-traumatic infall of gas that formed stars later. Whatever the
special machinery is which produces counterrotation or polar rings
instead of a co-planar, co-rotating distribution of gas and stars,
it seems not to be connected to the present galaxy density of their
environments.
 }

\maketitle

\section{Introduction}
In a previous paper \citep[ Paper I]{brocca} we studied the
environment of galaxies circled by a polar ring of gas and stars, a
kind of stellar system whose origin is still under discussion: in
recent theoretical works the polar rings are obtained by
means of a merging of galaxies \citep{bekki}, whereas in other studies,
it is ascribed either to the accretion of matter during a close
encounter with a nearby donor galaxy \citep{reshetnikov}, or to the 
capture of gas or stars from the environment \citep{whitmore,tremaine}. 
It is evident that in the three suggested scenarios, the present
environment of these galaxies should appear different in the richness 
of small satellites or bright companions. For instance, in the 
hypothesis of a close encounter, the donor galaxy should not be too
far from the polar ring galaxy, whereas in the case of diffuse gas,
no visible differences are expected. The epoch of the `second event'
may also play a role in the possibility of detecting the donor galaxy.
From the data of Paper I, we concluded that the environment of polar ring 
galaxies appears to be similar to that of normal
galaxies in relation to the richness, density or concentration of
satellites. We deduced that this result favours the infall of matter or
a very early merger theory.

A different result has been found by analysing the environment of other 
galaxy categories, whose peculiar origin has been connected in the 
literature to galactic interactions or mergers
\citep{gunn}, similar to polar rings.  In particular, the
environment of Seyfert galaxies appears to be richer in physical
companions than that of normal galaxies \citep{dahari, rafanelli}.
Similarly, the local galaxy density of 47 radio-loud elliptical and
lenticular galaxies appears higher than that of radio-quiet galaxies
by a factor 2-3 \citep{heckman}.  Even radio-loud QSO with
0.9$<$z$<$1.5 appear surrounded by a statistically significant excess
of galaxies \citep{hintzen}. Among these peculiar environments, the
apparent normality of polar ring galaxies looks anomalous and
distinguishes them from the other categories of astronomical objects
whose peculiarity or activity is ascribed to interaction or merger.
Therefore, we wish to see if other types of peculiar galaxies have an
environment similar to that of the active galaxies or if they are
`normal' in this respect, as the polar ring galaxies appear to be.

In this paper, the environment of another category of peculiar
galaxies, in which the rotation of the gas or of the stars is opposite
to that of most of the stars in the galaxy, is analysed. This
phenomenon, known as `counterrotation', presents a variety of aspects:
it may be present in the gas only \citep{bettoni84, caldwell, gall87,
ciri}, in a portion of the stars \citep{bender, franx, jed, bettoni89,
rubin, merrifield, prada, prada2} or in both
\citep{bertola, gall87}. The existing observations suggest that
counterrotation is a phenomenon which is present all along the Hubble sequence.
As in the case of polar ring galaxies, the origin of the
counterrotation has been attributed to different mechanisms. One of
these is the collision between the accretor galaxy and a small
satellite \citep{kennicut,thakar}. The difference to the formation
of a polar ring may reside in the satellite orbit, coplanar with the
disk of the accretor (and retrograde) to generate counterrotation
\citep{thakar}.  Another mechanism may be the merger of two
spirals of unequal mass \citep{balcells, bekki}, which is able to
transfer matter  as large as 10$^8$ \Msun or more in counterrotation
\citep{bettoni91, rubin, ciri, bettoni99}. A third alternative
involves an extended period of star- or gas- infall
during which the spin of the accreted matter changes rapidly
\citep{voglis, quinn2, merrifield, ostriker, rix}.

If counterrotation arises from accretion of a satellite galaxy, there
should be some peculiarity in the environment of these galaxies that
may be detectable in the present epoch, as happens in Seyfert or
radio-loud galaxies. Possibilities include: 1) a local
over-density of galaxies or 2)a larger number of satellites if the
accretion event occurred recently. A trace of the satellite accretion
may also remain if the phenomenon is not recent; for instance, a
deficit of satellites due to depletion of the environment may
differentiate them from a normal population. Finally, if the
counterrotation is generated by pure gas, pure star infall or by a
merger in the early epoch of galaxy formation, no traces of
differences should exist in the present surrounding field.

\section{Selection of the samples}

The first step in this work was to select a sample of galaxies with
counterrotation and, as wide as possible, a comparison sample of 'normal' 
galaxies. 
The latter should be representative of galaxies without
counterrotation but with distributions of luminosity and morphological
type similar to that of the galaxies with counterrotation. This
selection required a short analysis of the samples.

\subsection{Galaxies with counterrotation}

Our initial selection from the literature included all known examples
of gas or stellar counterrotation. For this selection the compilations
of \citet{gall96} and \citet{corsini}were used. We removed from
the lists a few doubtful examples and those with still unpublished
references.  These included NGC 2217, with a polar ring seen almost
face-on \citep{bettoni90} and NGC 4684, with gas streams along a
pole-on bar \citep{bettoni93} that mimic gas counterrotation. At the
end of this selection, our working list contained 49 galaxies, whose
names are indicated in the first column of Table \ref{cr}. They have
been divided according to the kind of counterrotation present. Four
systems show both kinds of counterrotation and have been considered
separately in the statistical analysis. They include IC 4889=IC 4891,
NGC 3593, NGC 4550 and NGC 7079. In the following, the label `gas cr'
or `stars cr' refer only to the pure cases of gas and star
counterrotation, while `all cr' or `cr galaxies' indicates both
samples plus these four galaxies.

The mean astrophysical parameters of these galaxies have been obtained
from the Lyon-Meudon Extragalactic Database, \citep{paturel}. They are
apparent and absolute magnitudes, radial velocity and morphological
type.

\subsection{Selection of the comparison galaxies}\label{KStxt}

In order to prepare a comparison sample, we made an initial list of
galaxies for which both gas and star rotation curves were
published. The principal source for this list was the Catalogue of
Spatially Resolved Kinematics of Galaxies \citep{hypercat} that is
available on the web. The published rotation curves for
each galaxy in the list was analysed, in order to determine the data-supported
co-rotation for the stars and gas. The comparison sample also includes
galaxies studied by \citet{kuijken}, where the authors claim that no
counter-rotating cores have been detected.

This list contains a high percentage of spirals, whereas galaxies with
counterrotation mostly have morphological types earlier than Sa. This
may alter the comparison between samples, with spiral galaxies
generally being present in a lower density environment.  To minimize
this difference, we extracted from the initial list of comparison
galaxies a sub-sample with a distribution of morphological types
similar to that of galaxies with counterrotation. This sub-sample was
checked to exclude the presence of other biases in their general
properties, applying a Kolgomorov-Smirnov test to their distributions
of absolute magnitudes $M_B$, red-shift and morphological type.  To
this end, we determined the cumulative frequency distribution $S(X)$
for each sample of observations by using the same interval for both
distributions. $S(X)$ is the fraction of data observed equal to or
less than $X$. Then, for each interval we subtracted one step function
from the other, evaluating D$_\alpha$=max$|S_1(X)-S_2(X)|$ which is
the maximum absolute difference found for the two distributions.  The
probability of the two samples having the calculated D$_\alpha$ and
coming from different galaxy populations is estimated by means of the
theoretical significance level SL, which is a function of the sizes of
the samples and is tabulated in statistical books. Comparing various
sub-samples of comparison galaxies we found one whose distribution of
intrinsic properties is not significantly different, at a confidence
level of SL=95\%, from that of all the cr galaxies (see Table
\ref{groups}). All the SL values calculated on the basis of the
corresponding D$_\alpha$ are lower than this value.

\begin{table*}
\caption{Summary of the Kolgomorov-Smirnov tests made to choose a comparison 
sample of normal galaxies. As explained in the text, D$_\alpha$ is the maximum 
fractional difference observed between the two distributions, while SL is the
significance level at which the two distributions compared are different. 
\label{groups} }
\begin{tabular}{lrrlrrlrr}
\hline
 & \multicolumn{2}{c}{gas cr vs. no cr} & &
\multicolumn{2}{c}{star cr vs. no cr} & & \multicolumn{2}{c}{all cr vs. no cr} \\
\cline{2-3}  \cline{5-6} \cline{8-9} \\
Parameter & D$_\alpha$ & SL & & D$_\alpha$ & SL & & D$_\alpha$ & SL \\
\hline
Morphological Type &  0.344 &  92.3\% & & 0.220 & 57.1\% & & 0.240 & 84.2\% \\
M$_B$    &  0.244 & 60.7\% & &  0.152 & 13.5\% & & 0.170 & 44.2\% \\
Red-shift    &  0.314 & 86.6\% & &  0.193 & 40.0\% & & 0.222 & 77.1\% \\
\hline
N. members (Garcia {\it et al.} 1993)   &  0.175 & $<$10\% & &  0.157 & $<$10\% & & 0.133 & 
$<$10\%  \\
N. members (Tully 1987)  &  0.218 & 22.9\% & & 0.034 & $<$10\%  & & 0.118 & 
$<$10\%  \\
Group velocity dispersion  (Tully 1987) & 0.096 & $<$10\% & & 0.112 &  $<$10\% & 
&  0.058 & $<$10\% \\
\hline
\end{tabular}
\end{table*}

An independent test of the quality of this comparison sample
concerns the large scale environment in which these galaxies are
located with respect to that of the cr galaxies. We then checked each
galaxy for which kind of environment it belongs to: field,
small groups and large clusters.  This classification was made
for galaxies with red-shift $\le$3000 \kms by \citet{tully} who
indicates for each one the richness and the velocity dispersion of
the group to which it belongs. The group richness is also available 
for galaxies brighter than magnitude 14 and with red-shift lower than
5500 \kms \citep{garcia}. 

These data show that the galaxies with counterrotation are
present in all kinds of environments, discovered both in clusters as
rich as Virgo as well in regions of the sky apparently empty of
companions. The distributions of richness and group velocity
dispersion show a large spread of values, without any tendency for
clustering of data at particular values. As performed for the absolute
magnitudes and morphological types, a Kolgomorov-Smirnov test was
applied to the samples (see Table \ref{groups}) and indicates that the
differences between galaxies with counterrotations and the comparison
sample are not significant.

These tests defined the final list for the comparison galaxies that
was used in the following Sections as a reference sample of
normal galaxies.  It is shown in the first column of Table \ref{ng}
and contains 43 galaxies.

\section{Data production and analysis}

To study the properties of the visible environment of these galaxies,
we searched all the objects present in the sky around every galaxy,
starting from the optical image databases available in the
literature. Our research was focused along different lines:

1) a first search of faint objects in the close neighbourhood of each galaxy,

2) a second search of bright objects that may have encountered the
galaxy within the last billion years;

3) a comparison of the density of galaxies within 40 Mpc, taken from
the literature \citep{tully2}, taking into account the presence of the
sample galaxies in groups of different hierarchy. This test is an
extension of that discussed in Sect. \ref{KStxt}. 
  
In the following, the galaxy to be studied, located at the centre of
each field, will be referred to as the `central galaxy', whereas all the
objects present in the selected field will be called `nearby objects',
even if they are in the foreground or in the background with respect
to the central galaxy. Only when the red-shift difference between the
nearby object and the central galaxy is lower than a fixed value,
described subsequently, will the object be defined as a
`companion'.

\subsection{Search for faint objects}

The first search was performed by adopting a searching radius of 100 kpc
and looking at all the objects present within that radius.  The search
was performed by extracting the data from the APM Sky Catalogue,
available for Internet access from the Observatory of Edinburgh
\citep{irwing}.  It contains data extracted by scanning and
photometrically calibrating the B and R plates of the Palomar Sky
Survey and the ESO/SRC J survey. It lists all the objects present in
the plates over the brightness level of 24 \sbu\ for the blue plates
and 23 \sbu\ for the red plates. They correspond to a limiting apparent
magnitude B=21.5 and R=20.0. For each object present in a field
corresponding to 100 kpc at the distance of the central galaxy, we
extracted the following parameters: $\alpha$ and $\delta$ co-ordinates,
B and R apparent magnitudes, semi-major axis, ellipticity and P.A. of
the ellipse fitting the image. In APM, galaxies are distinguished from
stars by means of a comparison of their Point Spread Function with
that of an `average' stellar image.

For most of the objects in the field the true distance is unknown, so
we adopted the following method to discard background and foreground
objects: all the parameters extracted from APM were converted in
distance from the central galaxy (kpc), absolute magnitude and linear
size (kpc){\it as if} all the objects were at the same distance as the
central galaxy. We expected that many background galaxies would appear
with a linear size or magnitude too small to be real
companions. Similarly, eventual foreground galaxies would appear too
big.  The limits taken to keep a galaxy were from 2 to 50 kpc for the
size and from M$_B$=-14 to M$_B$=-23. These figures have been chosen
as typical mean size and luminosity of the galaxies, taken from the
Local Group members \citep{zombeck, sandage} and from the Revised
Shapley-Ames Catalog \citep{sandage}. A limit of this method is given
by the fact that galaxies intrinsically fainter than M$_B$=-14 do
exist, e.g. Leo I, whose absolute magnitude is M$_B$=-9.6
\citep{sandage}.  It is clear that a more relaxed limit, for instance
M$_B \le$ -9, may include all the possible dwarf galaxies in the
surroundings of the central galaxy but will surely fill the sample of
nearby objects with a large number of background galaxies.  After a
set of tests with nearby central galaxies, we chose to limit the sample
at M$_R$=-14, bearing in mind that for fainter limits the possibility
of contamination by background objects is higher than the chances of
excluding possible dwarfs. Based on the distances of the central
object, we computed that an object of M$_B$=-14 with apparent B
magnitudes should always appear brighter than the APM detection limits
both for counterrotation and normal galaxy fields.  For this reason,
we are confident that most of the faint objects around the sample
galaxies are present in our search.

A problem to be faced in using the APM data is that the large galaxies
present in the field, especially if belonging to late morphological
types, appear fragmented in a set of small extended objects, reducing
their contribution to the local population of galaxies extracted from
APM. For this reason, we had to look at all the fields by plotting the
position and size of the objects present around the central galaxies
and we compared this map with the image of the same field extracted
from Palomar or ESO/SRJ atlases. When a galaxy is not included in the APM
catalogue, we extracted its position and size from other catalogues
and inserted it in our files. At the end of this correction, our set of 
data contained 40 galaxies with counterrotation
and 38 comparison galaxies, some systems not being present in the APM
catalogue. There are 16 galaxies with pure gas counterrotation, whereas 20
systems exhibit pure stellar counterrotation.  Four galaxies have both
types of counterrotation.

\subsection{Search for bright companions}

A second search concerned the detection of bright galaxies that may
have been gas or star donors during a close encounter. In this case
the search area has been defined in a different way: assuming that a
companion galaxy exists which may have encountered the central galaxy 
in the past, its linear distance will be R= $\Delta$V $\cdot \
\Delta$t, $\Delta$V being the relative velocity in space and $\Delta$t
the time elapsed since the encounter. If we assume a typical maximum
value $\Delta$V=600 \kms and a maximum elapsed time of 1 Gyr, the
maximum projected angular distance between the two galaxies seen at a
distance $d$ will be R$_{max}$[arcmin]= 2110.8/$d$[Mpc], which was our
search radius. This corresponds to a linear distance of 0.61
Mpc. We then searched galaxies within a radius of 0.61 Mpc from the 
central galaxy and having red-shift difference cz lower than 600 \kms 
with respect to it. 

This time we used the NED database, extracting position, red-shift,
apparent magnitude and size of every listed galaxy lying on the sky
inside R$_{max}$ and having $\Delta V \le$ 600 \kms. 
These data were converted into distances from the
central galaxy, differences in radial velocity, absolute magnitudes
and linear sizes. In this search, only galaxies with known redshift
were included in the sample. The final set of data includes 47
galaxies with counterrotation (18 gas cr, 25 star cr, 4 mixed) and 42
galaxies of the comparison sample. The excluded objects have too wide 
a field (such as NGC 253, with 14\fdg6 of search field) or have no
published red-shift (e.g NGC 2612).

\subsection{Large scale environment}

In addition to our data, we also used density values present in the
literature and computed on a much wider scale. We extracted from the
Nearby Galaxies Catalog \citep{tully2} the values of $\rho_{xyz}$,
the density of galaxies brighter than -16 mag determined within 40 Mpc
using a grid of 0.5 Mpc.

To have a direct comparison with our data, a `galaxy density' from APM
and NED data was calculated. In the first case, the total number
of observed faint objects within 100 kpc has produced $\rho_{APM}$,
the projected density of faint galaxies in units of Mpc$^{-2}$. The area
of 1 Mpc$^2$ is merely an arbitrary choice to plot the data and
produce large numbers. In the second case, the NED data were used to
produce a $\rho_{NED}$, in galaxies/Mpc$^3$, corresponding to
the mean density of bright objects with known redshift present in a
sphere of volume  0.95 Mpc$^3$, whose radius is the distance covered at a
velocity of 600 \kms in 1 Gyr.

The values so determined from our data are listed in the last columns
of Tables \ref{cr} and \ref{ng} and are plotted in the two panels of
Fig. \ref{plot1} versus $\rho_{xyz}$.

\begin{table*}
\caption{Statistical parameters $\rho_{ij}$ and galaxy densities for galaxies with 
counterrotation \label{cr} }
\tabcolsep 0.1truecm
\begin{tabular}{lrrrrrrrrrrrrrrrrr}
\hline
Name  & \multicolumn{6}{c}{APM data} & & \multicolumn{6}{c}{NED data} &  & \multicolumn{3}{c}{galaxy densities}  \\
\cline{2-7} \cline{9-14}  \cline{16-18}  \\
 & $\rho_{00}$ & $\rho_{01}$ & $\rho_{10}$ & $\rho_{11}$ & $\rho_{22}$ & $\rho_{3,2.4}$ &  &
$\rho_{00}$ & $\rho_{01}$ & $\rho_{10}$  & $\rho_{11}$ & $\rho_{22}$ & $\rho_{3,2.4}$ & & 
$\rho_{xyz}$ &  $\rho_{NED}$ & $\rho_{APM}$ \\
 &	&	&	&	&	&	& &	&	&	&	&	&	& &	&	&	\\
\hline
\multicolumn{18}{c}{gas cr} \\
\hline
 &	&	&	&	&	&	& &	&	&	&	&	&	& &	&	&	\\
ESO263-G48 	&		&		&		&		&		&		& &	1	&	0.168	&	0.084	&	0.014	&	0.000	&	0.000	& &		&	1	&		\\
IC  2006 	&		&		&		&		&		&		& &	14	&	1.347	&	4.680	&	0.412	&	0.028	&	0.014	& &		&	15	&		\\
NGC  128 	&	3	&	0.264	&	15.538	&	1.220	&	0.568	&	1.328	& &	10	&	1.842	&	17.609	&	2.354	&	1.677	&	5.188	& &		&	11	&	100	\\
NGC  253 	&	0	&	0.000	&	0.000	&	0.000	&	0.000	&	0.000	& &		&		&		&		&		&		& &	0.22	&		&	0	\\
NGC  497 	&	5	&	0.363	&	6.457	&	0.483	&	0.061	&	0.034	& &	3	&	0.663	&	0.580	&	0.120	&	0.005	&	0.001	& &		&	3	&	167	\\
NGC 1052 	&	7	&	0.506	&	17.999	&	0.930	&	0.209	&	0.273	& &	9	&	1.408	&	4.744	&	0.696	&	0.123	&	0.076	& &	0.49	&	9	&	234	\\
NGC 1216 	&		&		&		&		&		&		& &	2	&	0.584	&	2.839	&	0.857	&	0.470	&	0.584	& &		&	2	&		\\
NGC 2768 	&	2	&	0.147	&	2.472	&	0.174	&	0.015	&	0.006	& &	2	&	0.239	&	0.731	&	0.076	&	0.003	&	0.000	& &	0.31	&	2	&	67	\\
NGC 3497 	&	9	&	0.389	&	15.892	&	0.667	&	0.077	&	0.059	& &	1	&	0.142	&	1.022	&	0.146	&	0.021	&	0.010	& &		&	1	&	300	\\
NGC 3626 	&	2	&	0.080	&	4.148	&	0.193	&	0.028	&	0.025	& &	10	&	0.984	&	5.632	&	0.518	&	0.051	&	0.022	& &	0.32	&	11	&	67	\\
NGC 3941 	&	1	&	0.041	&	1.650	&	0.067	&	0.005	&	0.002	& &	3	&	0.374	&	0.804	&	0.103	&	0.004	&	0.001	& &	0.29	&	3	&	34	\\
NGC 4379 	&	3	&	0.152	&	4.801	&	0.263	&	0.036	&	0.026	& &	34	&	2.964	&	10.198	&	0.927	&	0.055	&	0.014	& &	2.89	&	36	&	100	\\
NGC 4546 	&	2	&	0.071	&	6.439	&	0.209	&	0.023	&	0.021	& &	2	&	0.151	&	0.541	&	0.042	&	0.001	&	0.000	& &	0.27	&	2	&	67	\\
NGC 4826 	&	1	&	0.023	&	1.299	&	0.029	&	0.001	&	0.000	& &		&		&		&		&		&		& &	0.20	&		&	34	\\
NGC 5252 	&	5	&	0.214	&	7.941	&	0.358	&	0.031	&	0.018	& &	1	&	0.242	&	0.272	&	0.066	&	0.004	&	0.001	& &		&	1	&	167	\\
NGC 5354 	&	6	&	0.525	&	23.129	&	2.038	&	1.301	&	3.848	& &	15	&	2.801	&	12.884	&	2.455	&	1.076	&	1.519	& &		&	16	&	200	\\
NGC 5898 	&		&		&		&		&		&		& &	5	&	0.788	&	6.385	&	0.856	&	0.252	&	0.292	& &	0.23	&	5	&		\\
NGC 7007 	&	7	&	0.241	&	8.932	&	0.298	&	0.013	&	0.004	& &	0	&	0.000	&	0.000	&	0.000	&	0.000	&	0.000	& &	0.14	&	0	&	234	\\
NGC 7097 	&	8	&	0.245	&	11.739	&	0.385	&	0.026	&	0.014	& &	2	&	0.197	&	0.697	&	0.060	&	0.002	&	0.000	& &	0.26	&	2	&	267	\\
NGC 7332 	&	3	&	0.174	&	11.996	&	0.629	&	0.222	&	0.334	& &	1	&	0.136	&	3.310	&	0.451	&	0.204	&	0.304	& &	0.12	&	1	&	100	\\
 &	&	&	&	&	&	& &	&	&	&	&	&	& &	&	&	\\
\hline
\multicolumn{18}{c}{stars cr} \\
\hline
 &	&	&	&	&	&	& &	&	&	&	&	&	& &	&	&	\\
IC  1459 	&	9	&	0.586	&	15.543	&	0.936	&	0.205	&	0.186	& &	14	&	2.033	&	2.529	&	0.400	&	0.026	&	0.006	& &	0.28	&	15	&	300	\\
NGC  936 	&	3	&	0.199	&	5.944	&	0.332	&	0.042	&	0.025	& &	5	&	0.580	&	3.792	&	0.439	&	0.062	&	0.038	& &	0.24	&	5	&	100	\\
NGC 1439 	&		&		&		&		&		&		& &	12	&	1.469	&	3.114	&	0.387	&	0.017	&	0.003	& &	0.45	&	13	&		\\
NGC 1543 	&	0	&	0.000	&	0.000	&	0.000	&	0.000	&	0	& &	11	&	1.288	&	1.983	&	0.211	&	0.004	&	0.000	& &	0.95	&	12	&	0	\\
NGC 1574 	&	1	&	0.021	&	1.828	&	0.039	&	0.002	&	0.001	& &	20	&	2.376	&	4.221	&	0.520	&	0.022	&	0.003	& &	0.98	&	21	&	34	\\
NGC 1700 	&		&		&		&		&		&		& &	0	&	0.000	&	0.000	&	0.000	&	0.000	&	0.000	& &		&	0	&		\\
NGC 2841 	&	1	&	0.021	&	2.288	&	0.049	&	0.002	&	0.001	& &	3	&	0.104	&	1.912	&	0.062	&	0.002	&	0.000	& &	0.13	&	3	&	34	\\
NGC 2983 	&		&		&		&		&		&		& &	1	&	0.250	&	0.215	&	0.054	&	0.003	&	0.000	& &	0.16	&	1	&		\\
NGC 3608 	&	4	&	0.199	&	11.118	&	0.524	&	0.083	&	0.087	& &	12	&	0.988	&	11.485	&	1.283	&	0.534	&	0.869	& &	0.56	&	13	&	134	\\
NGC 4073 	&	9	&	0.592	&	12.717	&	0.805	&	0.104	&	0.06	& &	10	&	1.946	&	4.652	&	0.758	&	0.080	&	0.031	& &		&	11	&	300	\\
NGC 4138 	&	3	&	0.088	&	3.942	&	0.117	&	0.005	&	0.002	& &	14	&	1.494	&	5.162	&	0.562	&	0.043	&	0.014	& &	0.84	&	15	&	100	\\
NGC 4472 	&	6	&	0.180	&	11.165	&	0.345	&	0.023	&	0.013	& &	71	&	4.502	&	46.441	&	2.437	&	0.235	&	0.283	& &	3.31	&	75	&	200	\\
NGC 4477 	&	4	&	0.213	&	7.383	&	0.386	&	0.053	&	0.041	& &	39	&	3.283	&	13.461	&	1.222	&	0.106	&	0.053	& &	4.06	&	41	&	134	\\
NGC 4596 	&	2	&	0.064	&	2.632	&	0.086	&	0.004	&	0.002	& &	12	&	1.574	&	4.078	&	0.505	&	0.041	&	0.013	& &	2.87	&	13	&	67	\\
NGC 4643 	&	0	&	0.000	&	0.000	&	0.000	&	0.000	&	0	& &	18	&	2.700	&	3.436	&	0.522	&	0.030	&	0.006	& &	0.25	&	19	&	0	\\
NGC 4816 	&	6	&	0.328	&	9.335	&	0.528	&	0.069	&	0.05	& &	26	&	4.333	&	9.467	&	1.505	&	0.359	&	0.464	& &		&	27	&	200	\\
NGC 5005 	&	0	&	0.000	&	0.000	&	0.000	&	0.000	&	0	& &	8	&	0.796	&	1.919	&	0.210	&	0.015	&	0.003	& &	0.28	&	8	&	0	\\
NGC 5322 	&	7	&	0.219	&	11.397	&	0.337	&	0.020	&	0.011	& &	9	&	1.344	&	1.211	&	0.153	&	0.003	&	0.000	& &	0.43	&	9	&	234	\\
NGC 5728 	&	6	&	0.175	&	8.930	&	0.287	&	0.022	&	0.016	& &	0	&	0.000	&	0.000	&	0.000	&	0.000	&	0.000	& &	0.18	&	0	&	200	\\
NGC 6684 	&		&		&		&		&		&		& &	7	&	1.137	&	0.736	&	0.092	&	0.002	&	0.000	& &	0.21	&	7	&		\\
NGC 6701 	&	15	&	0.549	&	27.578	&	0.977	&	0.082	&	0.058	& &	0	&	0.000	&	0.000	&	0.000	&	0.000	&	0.000	& &		&	0	&	500	\\
NGC 7217 	&	0	&	0.000	&	0.000	&	0.000	&	0.000	&	0	& &	0	&	0.000	&	0.000	&	0.000	&	0.000	&	0.000	& &	0.15	&	0	&	0	\\
NGC 7331 	&		&		&		&		&		&		& &	3	&	0.203	&	1.089	&	0.066	&	0.002	&	0.000	& &	0.33	&	3	&		\\
NGC 7796 	&	6	&	0.415	&	10.085	&	0.718	&	0.132	&	0.113	& &	0	&	0.000	&	0.000	&	0.000	&	0.000	&	0.000	& &		&	0	&	200	\\
UGC 9922A	&	8	&	0.287	&	12.187	&	0.445	&	0.029	&	0.014	& &	1	&	0.043	&	4.005	&	0.174	&	0.030	&	0.034	& &		&	1	&	267	\\
 &	&	&	&	&	&	& &	&	&	&	&	&	& &	&	&	\\
\hline
\multicolumn{18}{c}{gas plus stars cr} \\
\hline
 &	&	&	&	&	&	& &	&	&	&	&	&	& &	&	&	\\
IC  4889 	&	10	&	0.410	&	16.651	&	0.680	&	0.060	&	0.033	& &	1	&	0.108	&	0.315	&	0.034	&	0.001	&	0.000	& &	0.13	&	1	&	334	\\
NGC 3593 	&	0	&	0.000	&	0.000	&	0.000	&	0.000	&	0	& &	8	&	1.159	&	3.301	&	0.592	&	0.093	&	0.036	& &	0.19	&	8	&	0	\\
NGC 4550 	&	2	&	0.080	&	7.225	&	0.348	&	0.095	&	0.161	& &	7	&	0.315	&	7.640	&	0.396	&	0.053	&	0.050	& &	2.97	&	7	&	67	\\
NGC 7079 	&	0	&	0.000	&	0.000	&	0.000	&	0.000	&	0	& &	1	&	0.166	&	0.241	&	0.040	&	0.002	&	0.000	& &	0.19	&	1	&	0	\\
 &	&	&	&	&	&	& &	&	&	&	&	&	& &	&	&	\\
\hline
\end{tabular}
\end{table*}

\begin{table*}
\caption{Statistical parameters $\rho_{ij}$ and galaxy densities for the galaxies of the comparison sample \label{ng} }
\tabcolsep 0.1truecm
\begin{tabular}{lrrrrrrrrrrrrrrrrr}
\hline
Name  & \multicolumn{6}{c}{APM data} &  &  \multicolumn{6}{c}{NED data} &  & 
\multicolumn{3}{c}{galaxy densities}  \\
\cline{2-7} \cline{9-14} \cline{16-18} \\
 & $\rho_{00}$ & $\rho_{01}$ & $\rho_{10}$ & $\rho_{11}$ & $\rho_{22}$ & $\rho_{3,2.4}$ & 
  & $\rho_{00}$ & $\rho_{01}$ & $\rho_{10}$ & $\rho_{11}$ & $\rho_{22}$ &
$\rho_{3,2.4}$ & & $\rho_{xyz}$ &  $\rho_{NED}$ & $\rho_{APM}$ \\
 &	&	&	&	&	&	& &	&	&	&	&	&	& &	&	&	\\
\hline
 &	&	&	&	&	&	& &	&	&	&	&	&	& &	&	&	\\
IC 5063  	&	9	&	0.267	&	20.083	&	0.625	&	0.055	&	0.041	& &	2	&	0.290	&	0.227	&	0.037	&	0.001	&	0.000	& &		&	2	&	300	\\
NGC  488 	&	7	&	0.380	&	11.232	&	0.599	&	0.056	&	0.032	& &	3	&	0.291	&	2.543	&	0.200	&	0.015	&	0.006	& &	0.28	&	3	&	234	\\
NGC  628 	&	0	&	0.000	&	0.000	&	0.000	&	0.000	&	0	& &	7	&	0.553	&	2.422	&	0.204	&	0.012	&	0.003	& &	0.18	&	7	&	0	\\
NGC 1023 	&	2	&	0.069	&	2.261	&	0.078	&	0.003	&	0.001	& &	9	&	0.625	&	17.799	&	0.833	&	0.346	&	1.328	& &	0.57	&	9	&	67	\\
NGC 1275 	&		&		&		&		&		&		& &	8	&	1.754	&	4.078	&	0.911	&	0.156	&	0.078	& &		&	8	&		\\
NGC 1291 	&	2	&	0.053	&	5.320	&	0.147	&	0.013	&	0.01	& &	3	&	0.270	&	0.910	&	0.091	&	0.004	&	0.001	& &	0.14	&	3	&	67	\\
NGC 1300 	&		&		&		&		&		&		& &	4	&	0.395	&	1.669	&	0.187	&	0.015	&	0.006	& &	0.71	&	4	&		\\
NGC 1566 	&	2	&	0.063	&	3.283	&	0.104	&	0.006	&	0.002	& &	12	&	2.160	&	2.320	&	0.407	&	0.017	&	0.002	& &	0.92	&	13	&	67	\\
NGC 2613 	&		&		&		&		&		&		& &	1	&	0.124	&	0.169	&	0.021	&	0.000	&	0.000	& &	0.15	&	1	&		\\
NGC 2787 	&	4	&	0.121	&	5.705	&	0.184	&	0.012	&	0.006	& &	3	&	0.173	&	0.750	&	0.038	&	0.001	&	0.000	& &	0.06	&	3	&	134	\\
NGC 2903 	&	2	&	0.052	&	2.081	&	0.054	&	0.002	&	0	& &	0	&	0.000	&	0.000	&	0.000	&	0.000	&	0.000	& &	0.12	&	0	&	67	\\
NGC 2974 	&	5	&	0.112	&	9.964	&	0.220	&	0.012	&	0.007	& &	0	&	0.000	&	0.000	&	0.000	&	0.000	&	0.000	& &	0.26	&	0	&	167	\\
NGC 3079 	&	3	&	0.109	&	6.278	&	0.251	&	0.028	&	0.023	& &	3	&	0.213	&	6.280	&	0.368	&	0.112	&	0.207	& &	0.29	&	3	&	100	\\
NGC 3190 	&	5	&	0.263	&	12.673	&	0.725	&	0.159	&	0.197	& &	12	&	1.025	&	12.844	&	1.642	&	0.887	&	1.891	& &	0.52	&	13	&	167	\\
NGC 3198 	&	1	&	0.033	&	1.078	&	0.036	&	0.001	&	0	& &	1	&	0.033	&	0.125	&	0.004	&	0.000	&	0.000	& &	0.15	&	1	&	34	\\
NGC 3489 	&	0	&	0.000	&	0.000	&	0.000	&	0.000	&	0	& &	11	&	1.076	&	2.174	&	0.219	&	0.006	&	0.001	& &	0.39	&	12	&	0	\\
NGC 3516 	&	2	&	0.075	&	3.821	&	0.144	&	0.012	&	0.007	& &	0	&	0.000	&	0.000	&	0.000	&	0.000	&	0.000	& &	0.19	&	0	&	67	\\
NGC 3607 	&	6	&	0.216	&	22.133	&	0.806	&	0.171	&	0.323	& &	14	&	0.976	&	23.177	&	1.574	&	0.720	&	2.310	& &	0.34	&	15	&	200	\\
NGC 3623 	&	1	&	0.142	&	1.504	&	0.213	&	0.045	&	0.031	& &	7	&	1.236	&	4.603	&	1.055	&	0.399	&	0.310	& &	0.44	&	7	&	34	\\
NGC 3898 	&	5	&	0.169	&	9.638	&	0.302	&	0.022	&	0.015	& &	13	&	1.145	&	4.210	&	0.374	&	0.038	&	0.026	& &	0.56	&	14	&	167	\\
NGC 3921 	&	20	&	0.778	&	32.854	&	1.393	&	0.175	&	0.173	& &	2	&	0.476	&	1.976	&	0.311	&	0.050	&	0.027	& &		&	2	&	667	\\
NGC 3945 	&	1	&	0.024	&	1.546	&	0.037	&	0.001	&	0	& &	4	&	0.450	&	1.059	&	0.090	&	0.002	&	0.000	& &	0.50	&	4	&	34	\\
NGC 3962 	&	2	&	0.049	&	2.826	&	0.069	&	0.002	&	0.001	& &	2	&	0.211	&	0.423	&	0.040	&	0.001	&	0.000	& &	0.32	&	2	&	67	\\
NGC 4026 	&	1	&	0.051	&	1.121	&	0.057	&	0.003	&	0.001	& &	22	&	2.451	&	14.406	&	1.146	&	0.354	&	0.833	& &	1.71	&	23	&	34	\\
NGC 4027 	&		&		&		&		&		&		& &	14	&	2.139	&	6.055	&	0.624	&	0.044	&	0.021	& &	0.61	&	15	&		\\
NGC 4036 	&	2	&	0.051	&	3.585	&	0.096	&	0.006	&	0.004	& &	5	&	0.699	&	1.262	&	0.163	&	0.007	&	0.001	& &	0.44	&	5	&	67	\\
NGC 4111 	&	8	&	0.344	&	19.254	&	0.754	&	0.093	&	0.08	& &	13	&	1.138	&	25.703	&	1.337	&	0.340	&	0.754	& &	1.09	&	14	&	267	\\
NGC 4350 	&	1	&	0.086	&	3.484	&	0.300	&	0.090	&	0.117	& &	25	&	2.773	&	8.731	&	1.093	&	0.260	&	0.328	& &	2.72	&	26	&	34	\\
NGC 4379 	&	2	&	0.111	&	3.630	&	0.215	&	0.034	&	0.026	& &	34	&	2.964	&	10.198	&	0.927	&	0.055	&	0.014	& &	2.89	&	36	&	67	\\
NGC 4450 	&	1	&	0.024	&	3.509	&	0.086	&	0.007	&	0.006	& &	9	&	1.083	&	2.464	&	0.285	&	0.012	&	0.002	& &	1.88	&	9	&	34	\\
NGC 4565 	&	5	&	0.188	&	6.516	&	0.251	&	0.017	&	0.008	& &	2	&	0.363	&	2.851	&	0.400	&	0.102	&	0.101	& &	1.00	&	2	&	167	\\
NGC 4579 	&	1	&	0.029	&	1.595	&	0.046	&	0.002	&	0.001	& &	35	&	3.215	&	8.425	&	0.849	&	0.047	&	0.012	& &	3.26	&	37	&	34	\\
NGC 4594 	&	1	&	0.022	&	1.543	&	0.034	&	0.001	&	0	& &	2	&	0.203	&	0.467	&	0.044	&	0.001	&	0.000	& &	0.32	&	2	&	34	\\
NGC 4736 	&	1	&	0.020	&	1.020	&	0.021	&	0.000	&	0	& &		&		&		&		&		&		& &	0.42	&		&	34	\\
NGC 4762 	&	2	&	0.139	&	4.719	&	0.324	&	0.074	&	0.071	& &	16	&	1.534	&	6.618	&	0.726	&	0.131	&	0.114	& &	2.65	&	17	&	67	\\
NGC 5018 	&	5	&	0.354	&	7.384	&	0.491	&	0.116	&	0.085	& &	3	&	0.871	&	2.590	&	0.696	&	0.256	&	0.250	& &	0.29	&	3	&	167	\\
NGC 5055 	&	1	&	0.045	&	1.562	&	0.070	&	0.005	&	0.002	& &	4	&	0.287	&	1.403	&	0.067	&	0.002	&	0.000	& &	0.40	&	4	&	34	\\
NGC 5084 	&	4	&	0.306	&	7.022	&	0.463	&	0.062	&	0.038	& &	7	&	0.950	&	3.186	&	0.399	&	0.035	&	0.013	& &	0.29	&	7	&	134	\\
NGC 5746 	&	4	&	0.098	&	5.349	&	0.132	&	0.005	&	0.002	& &	7	&	0.594	&	2.978	&	0.296	&	0.032	&	0.014	& &	0.83	&	7	&	134	\\
NGC 5846 	&	3	&	0.457	&	4.665	&	0.662	&	0.261	&	0.231	& &	14	&	1.862	&	28.351	&	1.264	&	0.443	&	2.149	& &	0.84	&	15	&	100	\\
NGC 5866 	&	5	&	0.196	&	8.878	&	0.337	&	0.024	&	0.012	& &	3	&	0.513	&	1.858	&	0.350	&	0.077	&	0.039	& &	0.24	&	3	&	167	\\
NGC 6868 	&	7	&	0.491	&	14.982	&	1.007	&	0.270	&	0.43	& &	10	&	1.882	&	2.565	&	0.519	&	0.041	&	0.008	& &	0.47	&	11	&	234	\\
NGC 7052 	&		&		&		&		&		&		& &	1	&	0.161	&	0.093	&	0.015	&	0.000	&	0.000	& &		&	1	&		\\
 &	&	&	&	&	&	& &	&	&	&	&	&	& &	&	&	\\
\hline
\end{tabular}
\end{table*}

\section{Statistical parameters}\label{sp}

The statistical analysis of all fields for the cr galaxies and
comparison sample was done defining a set of
density parameters for each field:

\begin{equation}
\rho_{ij}=\sum_k r_k^{-i} D_k^j
\end{equation}

where $r_k$ is the projected distance between the central galaxy and
the {\it k}th galaxy, $D_k$ is the projected diameter of the
{\it k}th galaxy and (i,j) assumes the values 0, 1, (2,2) and
(3,2.4). We normalised the D$_k$ and r$_k$ values in units of
100 kpc. 

The first three parameters describe the environment of the galaxies
with different criteria: the population ($\rho_{00}$), the total size
of sky covered by surrounding galaxies ($\rho_{01}$), the
concentration of surrounding galaxies $\rho_{10}$. The remaining three
are linked with these: $\rho_{11}$ is proportional to the
gravitational potential and $\rho_{22}$ to the gravitational force
exerted by the surrounding galaxies on the central object, whereas
$\rho_{3,2.4}$ is proportional to the tidal interaction between the
surrounding galaxies and the central one.  The last two parameters
amplify the effects present in the parameter $\rho_{11}$.  These
parameters were chosen in accordance with similar studies
\citep{heckman, fuentes}.

After conversion into linear units, as described in the precedent paragraphs,
the diameters and the distances from the central galaxies were
converted into units of 100 kpc. The resulting values of the $\rho_{ij}$
for the field of each central galaxy are listed in Table
\ref{cr} and Table \ref{ng}. Results from the APM
database and NED are presented together in these tables.

After defining the $\rho_{ij}$ parameters for the two samples of
counter-rotating galaxies and  normal galaxies, a
Kolgomorov-Smirnov test was applied to the $\rho_{ij}$
parameters of the local environment (APM data, galaxies within 100
kpc), to that of the intermediate environment (NED, bright galaxies
with similar red-shift) and to $\rho_{xyz}$ densities \citep[on 40 Mpc
scale]{tully2}.  The results are shown in Table \ref{KS}.

\section{Results}

\begin{figure}
\centering
\resizebox{\hsize}{!}{\includegraphics{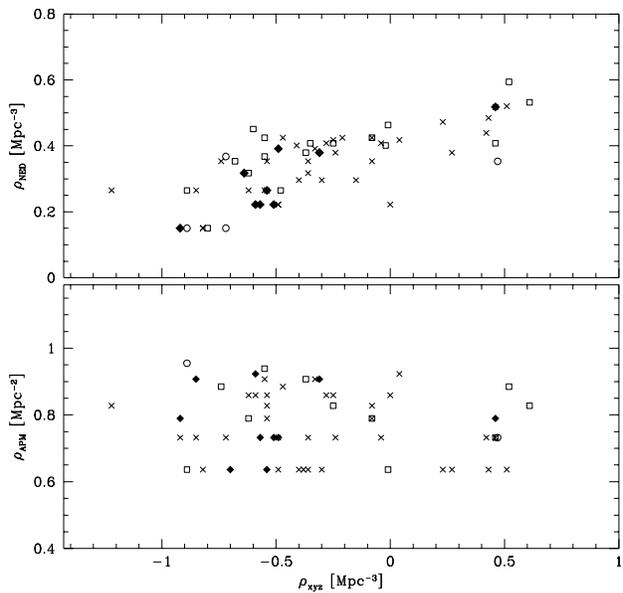}}
\caption{Plot of the density of objects around all the sample galaxies in 
environments with different size. The galaxies with pure gas
counterrotation or pure stellar counterrotation are indicated by full
diamonds and open squares respectively; the normal galaxies are
indicated by crosses. The galaxies with both gas and stellar
counterrotation are plotted with open circles. {\it Top panel}: plot
of ($\rho_{NED}$) the density of galaxies/Mpc$^3$ present in NED
database with a crossing time lower than 1 Gyr versus ($\rho_{xyz}$),
the density of galaxies computed within 40 Mpc from
\citet{tully2}.  {\it Bottom panel}: Plot of ($\rho_{APM}$), the density of 
galaxies extrapolated to a square of 1 Mpc side on the sky, versus 
($\rho_{xyz}$). These plots are discussed in the text.}
\label{plot1}
\end{figure}

We begins by considering the local densities
at three different scales: a close environment within 100 kpc, an
intermediate environment inside 0.61 Mpc, defined by a crossing time
of 1 Gyr, and a large scale environment inside 40 Mpc. These densities
were extracted from the APM, NED and \citet{tully2} catalogues
respectively. The APM data refers to a projected galaxy density
$\rho_{APM}$, whereas the NED and \citet{tully2} data use the red-shift
to define `volume' densities $\rho_{NED}$ and $\rho_{xyz}$
respectively.  When these densities are plotted against each other
(Fig. \ref{plot1}) we can see that $\rho_{xyz}$, and
$\rho_{NED}$ are correlated. In particular, excluding NGC 4379, all
the remaining galaxies with gas counterrotation belong to groups where
the density of galaxies $\rho_{xyz}<$ 0.5 galaxies Mpc$^{-3}$ and
$\rho_{NED}<$ 1.8 galaxies Mpc$^{-3}$. On the contrary, the projected 
densities extracted from APM data span the entire range of 
plotted values, without any particular clustering of points. This may
be affected by the presence of background objects, which alter the
sample, despite our selection criteria defined in Sect. \ref{sp}, or
may suggest that there is no particular clustering of objects in the
surrounding of our sample galaxies. However, bearing in mind that our
cut-off levels in magnitudes and sizes for possible companion galaxies
were quite high, it is more likely that a few satellites were missed
rather than having a significant contamination by background
objects. We think then that the plot reflects the real situation of
the densities existing within 100 kpc of our sample objects.

\begin{table}
\caption{Summary of Kolgomorov-Smirnov tests on the density parameters 
calculated from the data extracted from APM and NED
databases. D$_\alpha$ is the maximum difference observed between the
two distributions, whereas SL is the percentage significance level at
which the two distributions compared are different. In the last three
lines, we show the results of the tests applied to the density
$\rho_{xyx}$ and $\rho_{NED}$ (in galaxies/Mpc$^3$) and $\rho_{APM}$
(in galaxies/Mpc$^2$) described in the text.\label{KS} }

\tabcolsep 0.1truecm
\begin{tabular}{lrrlrrlrr}
\hline
 & \multicolumn{2}{c}{gas cr vs. no cr} & & \multicolumn{2}{c}{stars cr vs. no cr}
& & \multicolumn{2}{c}{all cr vs. no cr} \\
\cline{2-3}  \cline{5-6} \cline{8-9} \\
Param. & D$_\alpha$ & SL & & D$_\alpha$ & SL & & D$_\alpha$ & SL \\
\hline
\multicolumn{9}{l}{APM (100 kpc environment)} \\
\hline
$\rho_{00}$    &  0.178 & 13.3\% & &  0.292 & 78.6\% & & 0.217 & 64.9\% \\
$\rho_{01}$    &  0.283 & 66.9\% & &  0.258 & 64.9\% & & 0.233 & 72.9\% \\
$\rho_{10}$    &  0.257 & 54.9\% & &  0.237 & 54.3\% & & 0.207 & 58.8\% \\
$\rho_{11}$    &  0.207 & 28.5\% & &  0.179 & 20.9\% & & 0.179 & 40.6\% \\
$\rho_{22}$    &  0.145 & $<$10  \% & & 0.121 & $<$10  \% & & 0.096 & $<$10  \% \\
$\rho_{3,2.4}$ &  0.171 & 10.4\% & &  0.079 & $<$10  \% & & 0.050 & $<$10  \% \\
\hline
\multicolumn{9}{l}{NED (1 Gyr similar redshift companions)} \\
\hline
$\rho_{00}$    &  0.167 & 12.7\% & &  0.130 & $<$10  \% & & 0.056 & $<$10  \% \\
$\rho_{01}$    &  0.206 & 34.6\% & &  0.242 & 67.9\% & &  0.166 & 40.1\% \\
$\rho_{10}$    &  0.071 & $<$10  \% & & 0.129 & $<$10  \% & & 0.058 & $<$10  \% \\
$\rho_{11}$    &  0.135 & $<$10  \% & & 0.129 & $<$10  \% & & 0.095 & $<$10  \% \\
$\rho_{22}$    &  0.111 & $<$10  \% & & 0.150 & 12.8\% & &  0.097 & $<$10  \% \\
$\rho_{3,2.4}$ &  0.056 & $<$10  \% & & 0.154 & 15.4\% & &  0.097 & $<$10  \% \\
\hline
\multicolumn{9}{l}{densities of galaxies from different catalogues} \\
\hline
$\rho_{xyz}$  &  0.372 & 84.2\% & &  0.140 & $<$10  \% & & 0.238 & 71.6\% \\
$\rho_{NED}$  &  0.206 & 34.6\% & &  0.155 & 15.9\% & & 0.063 & 20.4\% \\
$\rho_{APM}$  &  0.178 & 13.3\% & &  0.203 & 34.8\% & & 0.153 & 22.3\% \\
\hline
\end{tabular}
\end{table}
 
This segregation/concentration for galaxies with gas counterrotation
in Fig. \ref{plot1} is confirmed by a Kol\-gomorov--Smirnov test
applied to the galaxy densities around the whole sample, except for
NGC 4379. The population of galaxies with gas counterrotation and the
population of normal galaxies appear to be different at a significance
level of 93.7\%. However, there is no {\it a priori} justification for
the exclusion of NGC 4379. Its inclusion reduces the significance
level to 84\% and weakens the difference between populations. The same
test applied to the NED and APM densities indicates that the
difference between samples decreases when smaller environments are
considered (see the last lines of Table \ref{KS}). We may conclude
that, in the limited number of galaxies with pure gas counterrotation
available, they tend to lie in less dense groups, on scales larger
than $\sim$0.5 Mpc.

Looking at the other studied parameters, described by the $\rho_{ij}$
quantities (Tables \ref{cr} and \ref{ng}), Kolgomorov-Smirnov tests
indicate that no marked differences are evident in APM or in the NED
data. In fact, no significance level is above $\sim$80\% (Table
\ref{KS}).

\section{Conclusions}

We deduce that in general the surrounding regions of galaxies with
counterrotation do not appear {\it statistically} different from those
of normal galaxies. This result is similar to that found in Paper I
for the environment of polar ring galaxies but distinguish our
galaxies from the other active galaxy categories \citep{dahari,
heckman, hintzen, rafanelli}.

Among the hypotheses presented in the Introduction about the origin of
counterrotation and polar rings, our result tend to disprove that of a
recent interaction with a small satellite or a galaxy with similar
size. If such a process is at the origin of the counterrotation
phenomenon \citep{balcells, bekki, kennicut,thakar}, it cannot be
younger than 1 Gyr, the crossing time for the volumes of space studied
in this paper. Otherwise it is difficult to conceive that no trace of
the donor galaxy remains in the surrounding space, both as a
single galaxy present in the NED archive or in a form detectable in
APM data as diffuse surrounding objects.

This result, to a first approximation, support the hypothesis
that all galaxies are born from a merger process of
smaller objects occurring early in their life. However,
only a few galaxies that we know of develop counterrotation and polar
rings. It may be natural to attribute this peculiarity to a richer
environment, which makes the possibility of collisions easier. Our data
are also contrary to this hypothesis, because the environment of such 
galaxies does not appear to be richer in satellites. This is different to
Seyfert or radio-loud galaxies which lie in environments with a higher
density of companions.  On the contrary, if a weak tendency exists 
for galaxies with gas counterrotation only, it is seen
in regions of space where the large scale density of
galaxies is smaller. Whatever the special machinery is which produces
counterrotation or polar rings instead of a co-planar, co-rotating
distribution of gas and stars, it is not connected to the present
galaxy density of their environments.

An alternative mechanism to form counterrotation and polar rings may
arise from a continuous, non traumatic infall of gas that later formed
stars \citep{voglis, quinn2, merrifield, ostriker, rix}. In such a
case the past and present visible environment of these galaxies would
appear similar to that of the other galaxies, even if the
process is still active.  This explanation is consistent with all the results
we found. The slow infall of matter on a galaxy should not alter
either its luminosity distribution or its stellar kinematics, untill
the accreted mass is large enough to generate tidal actions. The
galaxies with counterrotation and the polar rings may in such a scenario
appear relaxed or in equilibrium, even if some star formation is active
(see the polar rings of NGC 4650A and NGC 5128).

The study of the peculiar galaxies that present gas accretion is still
under discussion of the possible models to explain their origin and
evolution. It is currently impossible to decide between the previous,
perhaps incompatible, theories (early merging with special dynamical
conditions or continuous slow infall). On the other hand, the
observations of galaxies with counterrotation, begun in 1984, provide
clues to their evolution, but are still not conclusive. To solve the
problem, we are planning to study the gas content of galaxies with
counterrotation and polar rings
\citep{bettoni01}.

Acknowledgments

We would like to thank Dr. J. Sulentic for useful suggestions regarding
this paper.  This research was done using the LEDA database,(
leda.univ-lyon1.fr), of the NASA/IPAC Extragalactic Database (NED,
nedwww.ipac.caltech.edu) which is operated by the Jet Propulsion
Laboratory, California Institute of Technology, under contract to
the National Aeronautics and Space Administration and of the Hypercat
database (www-obs.univ-lyon1.fr/hypercat). This search has been granted
by the funds 60\%-2000 of the Universit\`a di Padova.

\end{document}